\documentclass[10pt,journal,comsoc]{IEEEtran} 
\usepackage{multirow}
\usepackage[textwidth=1cm,backgroundcolor=yellow,textsize=tiny]{todonotes}

 \usepackage{amsmath,amsfonts, amsthm, bm}

\usepackage{comment, tagging}

\usetag{long} % JUST FOR THE LONG VERSION
\usepackage{colortbl}
\usepackage{stackengine} % Allows having sub-floats without (a), (b), ...

\usepackage{array}
\newcolumntype{P}[1]{>{\centering\arraybackslash}p{#1}}
\newcolumntype{M}[1]{>{\centering\arraybackslash}m{#1}}
\usepackage{setspace}

\usepackage[compress]{cite}
\usepackage{tabto}
\usepackage[noend]{algpseudocode}
\usepackage{algorithm}
\usepackage{varwidth} % Used for indenting broken long line in PGM's pseudo-code
\usepackage{xspace, paralist}
\usepackage{url}
\PassOptionsToPackage{hyphens}{url}
\usepackage{float}
\usepackage{graphicx, pstricks, pst-node,pst-plot}
\usepackage{subfloat, tabu, color, subfiles}
\usepackage[caption=false,labelfont=sf,textfont=sf]{subfig}
\usepackage{textcomp, stfloats, verbatim}

\algnewcommand{\AND}{\textbf{and}\xspace}
\algnewcommand{\OR}{\textbf{or}\xspace}
\algnewcommand\algorithmicforeach{\textbf{for each}}
\algdef{S}[FOR]{ForEach}[1]{\algorithmicforeach\ #1\ \algorithmicdo}

\algrenewcommand\algorithmicindent{1.0em}%

\makeatletter
\algnewcommand{\LineComment}[1]{\Statex \hskip\ALG@thistlm \(\triangleright\) #1}
\makeatother

\newbox\statebox
\newcommand{\myState}[1]{%
    \setbox\statebox=\vbox{#1}%
    \edef\thealgruleheight{\dimexpr \the\ht\statebox+1pt\relax}%
    \edef\thealgruledepth{\dimexpr \the\dp\statebox+1pt\relax}%
    \ifdim\thealgruleheight<.75\baselineskip
        \def\thealgruleheight{\dimexpr .75\baselineskip+1pt\relax}%
    \fi
    \ifdim\thealgruledepth<.25\baselineskip
        \def\thealgruledepth{\dimexpr .25\baselineskip+1pt\relax}%
    \fi
    \State #1%
    \def\thealgruleheight{\dimexpr .75\baselineskip+1pt\relax}%
    \def\thealgruledepth{\dimexpr .25\baselineskip+1pt\relax}%
}
\newcommand{\abs}[1]{\left\vert#1\right\vert}

\usepackage{mathtools, nccmath}

\DeclarePairedDelimiter{\nint}\lfloor\rceil

\usepackage{stackengine}

\usepackage[breaklinks=true]{hyperref}
\newcommand{\hyperSize}{\mathbf{H}}
\newcommand{\hyperBit}{h}
\newcommand{\hyperVec}{\Vec{\hyperBit}}
\newcommand{\hyperVal}{\mathbf{E}}
\newcommand{\expSize}{\hyperVal}

\newcommand{\expMaxSize}{\hyperVal_{max}}
\newcommand{\val}{V}
\newcommand{\valMax}{\val_{max}}
\newcommand{\expBit}{e}
\newcommand{\expVec}{\Vec{\expBit}}
\newcommand{\expVal}{E}
\newcommand{\expMinVal}{\expVal_{min}}
\newcommand{\bias}{B}
\newcommand{\mantBit}{m}
\newcommand{\mantVec}{\Vec{\mantBit}}
\newcommand{\mantSize}{\mathbf{M}}
\newcommand{\mantMinSize}{\mantSize_{min}}
\newcommand{\mantVal}{M}
\newcommand{\cntrVec}{\vec{n}}
\newcommand{\cntrSize}{\mathbf{N}}
\newcommand{\cntrVal}{N}

\DeclareMathOperator{\ffp}{F2P}
\DeclareMathOperator{\sr}{SR}
\DeclareMathOperator{\lr}{LR}
\DeclareMathOperator{\si}{SI}
\DeclareMathOperator{\li}{LI}
\newcommand{\expValSr}{$\expVal_{\sr}$}
\newcommand{\expValLr}{$\expVal_{\lr}$}
\newcommand{\ffpSr}{$\ffp_{\sr}$}
\newcommand{\ffpLr}{$\ffp_{\lr}$}
\newcommand{\ffpSi}{$\ffp_{\si}$}
\newcommand{\ffpLi}{$\ffp_{\li}$}

\newcommand{\ffpSrOne}{$\ffp_{\sr}^1$}
\newcommand{\ffpSrTwo}{$\ffp_{\sr}^2$}
\newcommand{\ffpLrOne}{$\ffp_{\lr}^1$}
\newcommand{\ffpLrTwo}{$\ffp_{\lr}^2$}
\newcommand{\ffpSiOne}{$\ffp_{\si}^1$}
\newcommand{\ffpSiTwo}{$\ffp_{\si}^2$}
\newcommand{\ffpLiOne}{$\ffp_{\li}^1$}
\newcommand{\ffpLiTwo}{$\ffp_{\li}^2$}

\date{}
\title{Floating-floating point: a highly accurate number representation with flexible Counting ranges
}

\author{Itamar Cohen,~\IEEEmembership{Member,~IEEE,} and
Gil Einziger~\IEEEmembership{Member,~IEEE,} 
\thanks{I. Cohen is with 
the Department of Computer Science,  
Ariel University, Ariel 40700, Israel; e-mail: itamarc@ariel.ac.il.
Gil Einziger with 
the Department of Computer Science, 
Ben-Gurion University of the Negev, Beer Sheva 8410501, Israel; 
e-mail: gilein@bgu.ac.il.} 
}

\begin{document}
\maketitle
\begin{abstract}
Efficient number representation is essential for federated learning, natural language processing, and network measurement solutions. 
Due to timing, area, and power constraints, such applications use narrow bit-width (e.g., 8-bit) number systems. The widely used floating-point systems exhibit a trade-off between the counting range and accuracy. 
This paper introduces Floating-Floating-Point (F2P) -- a floating point number that varies the partition between mantissa and exponent. Such flexibility leads to a large counting range combined with improved accuracy over a selected sub-range. Our evaluation demonstrates that moving to F2P from the state-of-the-art improves network measurement accuracy and federated learning. 
\end{abstract}


\section{Introduction}\label{sec:intro}

\IEEEPARstart{T}he double-precision Floating-Point (FP64) number representation provides an excellent counting range and accuracy. However, these benefits come with power, area, and delay overheads, which are unacceptable for arithmetic-intensive tasks such as inference in neural networks~\cite{Quant_white_paper, FP8_quant}, network telemetry~\cite{Sketch_survey_arxiv20}, and computer vision~\cite{Pixel_cntrs17}.
Consequently, such applications commonly use shorter number representations, such as INT8~\cite{Quant_white_paper}, FP8~\cite{FP8_quant}, FP16, BFloat16~\cite{BFloatBF16GoogleCloud}, and TensorFloat32~\cite{TensorFloat32Nvidia}.

Each number representation offers some trade-off between counting range and accuracy. For instance, an $\cntrSize$-bits integer provides a fixed resolution of 1 and a counting range of $2^\cntrSize$. 
Floating-point-based systems provide a more flexible resolution, by partitioning the number into a sign bit, $\expSize$ exponent bits, and $\cntrSize-1-\expSize$ mantissa bits. To increase the counting range, one may enlarge the exponent field $\expSize$ -- but this comes at the cost of shrinking the mantissa field, which results in deteriorating the accuracy throughout the counting range~\cite{FP8_quant}. 
Dynamic SEAD~\cite{Sead_ToN} broke this trade-off between counting range and accuracy by introducing a scalable exponent field, whose size is unary-encoded. However, the unary encoding is space-inefficient, leaving only very little space for the mantissa, which results in very low accuracy.

This paper presents Floating-Floating-Point (F2P) -- a novel number system that varies the number of exponent bits. As a result, F2P provides both an extensive counting range and high accuracy at a chosen part of the range. 
For instance, F2P can focus on accurately representing small numbers, which is a desired property in neural networks that are characterized by short-tail distribution~\cite{FP8_quant}. On the other hand, if desired, F2P can be customized to 
favor the accuracy of large numbers, as desired by other neural network models~\cite{FP8_quant} and by typical network telemetry applications~\cite{Sketch_survey_arxiv20}.

\begin{figure}[t!]
    \centering
    \includegraphics[width=\columnwidth]{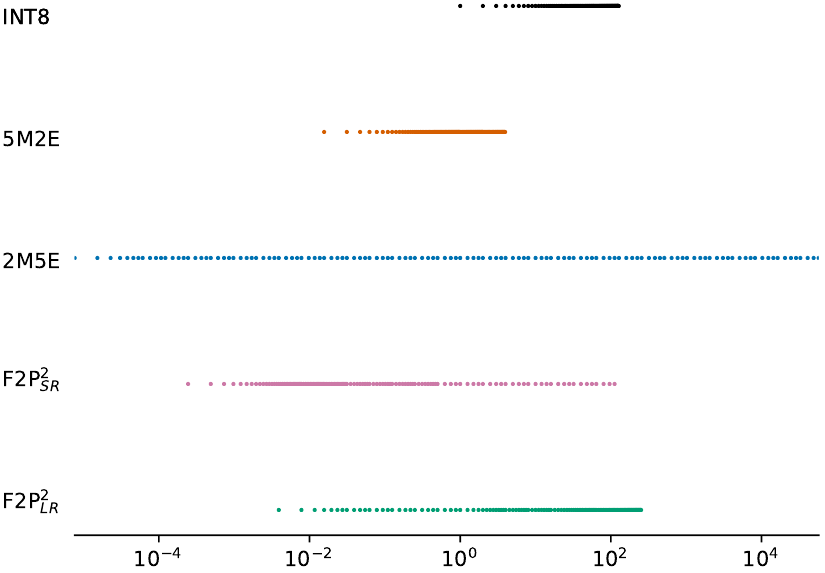}
    \caption{The positive values that are represented by different 8-bit grids. Existing solutions provide either good resolution and a small counting range (INT8, 5M2E) or a large counting range with inferior resolution (2M5E). Our F2P flavors (pink and green) combine a relatively large range with a dense representation for a chosen sub-range.}
    \label{fig:grids}
\end{figure}

To gain some intuition about F2P benefits, consider Fig.~\ref{fig:grids}, which depicts the positive values representable by different 8-bit grids. 
To capture the versatile counting ranges, the X-axis is logarithmic. xMyE denotes an FP representation with x mantissa bits and y exponent bits.
INT8 and 5M2E obtain high accuracy (captured by dense points distribution), but their counting ranges are limited.
2M5E, on the other extreme, stretches over a wide counting range but with inferior accuracy throughout.
\ffpSrTwo and \ffpLrTwo, two flavors of F2P (to be detailed later), combine the advantages of 5M2E and 2M5E by covering large counting ranges while still accurately representing a significant sub-range. Such a grid is very desired for applications that have to deal with values drawn from large counting ranges while focusing on the accuracy of sub-range, e.g., large values (``elephants'') in network telemetry~\cite{Sketch_survey_arxiv20}, or outliers in neural networks~\cite{FP8_quant}.

\paragraph*{Our contribution} In this paper, we introduce F2P -- a novel number system with flexible exponent and mantissa fields. We develop several flavors of F2P, customized to boost the accuracy in different ranges. Furthermore, some of F2P's flavors represent only integer values, a desired property in numerous applications, such as counting flow size and flow volume~\cite{Sketch_survey_arxiv20}.
To ease the adoption of F2P in practical systems, our design keeps most of the mechanisms and formulas used in standard FP systems.
Our evaluation demonstrates multiple scenarios in which F2P outperforms state-of-the-art solutions.

\section{The F2P Number System}\label{sec:F2P}
\noindent This section describes the Floating Floating-Point (F2P) number representation.
We begin with some preliminaries. Then, we outline the classical representation of floating point (FP) numbers. Next, we provide a high-level description of our suggested F2P number representation and then turn to specify its details.
% and analyze its properties.

\subsection{Preliminaries}

For ease of presentation, we focus on representing unsigned (non-negative) numbers in our technical description. 
However, F2P can represent signed numbers by defining the  MSB as the sign bit. 

We view a number as an $\cntrSize$-bit vector, $\cntrVec = \cntrVec_{\cntrSize-1} \dots \cntrVec_0$. Let $\cntrVal(\cntrVec)$ denote the value represented by $\cntrVec$.
A number $\cntrVec$ is partitioned into several fields. The {\em exponent} field consists of an $\expSize$-bits exponent vector, $\expVec = \expBit_{\expSize-1} \dots \expBit_{0}$. 
Let $\expMinVal$ denote the minimal possible value encoded by the exponent vector. Using a standard unsigned binary encoding (as done in the FP system),  $\expMinVal=0$. However, we consider a different encoding scheme, implying that $\expMinVal$ is not necessarily zero.
The {\em mantissa} field consists of an $\mantSize$-bits vector $\mantVec = \mantBit_{-1} \dots \mantBit_{-\mantSize}$.
Following previous work~\cite{CEDAR_Tsidon, Sead_ToN, Morris1978, FP8_quant, Quant_white_paper}, we discard the representation of special values (e.g., infinity, underflow), as these values are typically irrelevant or unnecessary in applications that use short-width number systems.

We use the ``empty sum is zero'' rule to simplify expressions. For instance, we assign $\sum_{i=1}^0 x_i=0$ regardless of the actual values of the $x_i$'s.
For ease of reference, our main notations are summarized in Tab.~\ref{tab:notations}.

\subsection{Floating Point Number Representation}\label{sec:FP}
\begin{table}[t!]
    \centering
    \footnotesize
    \caption{Main notation}\label{tab:notations}
	\begin{tabular}{|c|l|}   
         \hline
         Symbol & Description \\
         \hline \hline 
         $\cntrSize$ & number of bits in the number\\ \hline
         $\cntrVec$ & Number as a vector: $\cntrVec = n_{\cntrSize-1} \dots n_0$ \\ \hline
         $\cntrVal$ & The value represented by the vector $\cntrVec$ \\ \hline 
         $\expSize$ & number of bits in the exponent field \\ \hline
         $\expVec$ & Exponent field's content: $\expVec = \expBit_{\expSize-1} \dots \expBit_{0}$\\ \hline
         $\expVal$ & Value of the exponent (Eq.~\ref{eq:def_expVal_mantVal}) \\ \hline
         $\expMinVal$ & Minimum value encoded by the exponent\\ \hline
         $\mantSize$ & \# of bits in the mantissa field\\ \hline         
         $\mantMinSize$ & minimum \# of bits in the mantissa field\\ \hline
         $\mantVec$ & Mantissa field's content: $\mantVec = m_{-1} \dots m_{-\mantSize}$\\ \hline
         $\mantVal$ & Value of the mantissa (Eq.~\ref{eq:def_expVal_mantVal}) \\ \hline
         $\bias$ & Bias factor added to the exponent (Eq.~\ref{eq:FP_val}) \\ \hline
         \rule[1pt]{0pt}{2.0ex}    
          F2P$^\hyperSize$ & An F2P system with $\hyperSize$ hyper-exp bits\\ \hline
          % $\hyperSize$ & Size of the hyper-exp field\\ \hline
         $\hyperVec$ & Hyper-exp field's content: $\hyperVec = \hyperBit_{\hyperSize-1} \hyperBit_{\hyperSize-2} \dots h_0$\\ \hline
         $\valMax$ & \# of distinct exponent values of (Eq.~\ref{eq:valMax}) \\ \hline
       %  $f$ & Offset added to the number \\ \hline
         % $\cntrValSet$ & Set of values that the counter can represent\\ \hline
         % $\cntrValMax$ & Max value that the counter can represent. \\
         % \hline 
         % $\targetVal$ & Target value after modification\\ \hline 
         % $\targetValLB$ & See~\eqref{eq:def_targetValLB_UB}\\ \hline 
         % $\targetValUB$ & See~\eqref{eq:def_targetValLB_UB}\\ \hline 
    \end{tabular}
\end{table}
The FP number system associates the exponent and the mantissa vectors with the values
\begin{align}\label{eq:def_expVal_mantVal}
\expVal(\expVec) = \sum_{i=0}^{\expSize-1} \expBit_i \cdot 2^i, \quad \quad
\mantVal(\mantVec)= \sum\limits_{i=-\mantSize}^{-1} \mantBit_i \cdot 2^i. 
\end{align}

Using~\eqref{eq:def_expVal_mantVal}, the minimal possible value of the exponent is $\expMinVal=0$.
To tune the counting range, the floating-point system adds to the exponent a {\em bias} $\bias$, which we shortly discuss. Given the values of the exponent $\expVal$ and of the mantissa $\mantVal$, the value of the number is

\begin{align}\label{eq:FP_val}
    \cntrVal (\expVal, \mantVal) = 
    \begin{cases}
    2^{\expVal+\bias} \left( 1 + \mantVal \right) & \expVal > \expMinVal \\
    2^{\expVal +\bias + 1} \cdot \mantVal  & \expVal = \expMinVal    
    \end{cases}
\end{align}

FP's bias is commonly determined based on the following principle.

\noindent{\em The symmetrical power principle.} FP systems typically aim to make the minimal and maximal powers of 2 that may appear in Eq.~\ref{eq:FP_val} symmetrical w.r.t zero. Hence, the bias is commonly set to $\bias = -2^{\expSize-1}$, implying that the minimal, and maximal values of $\expVal+\bias$ are $-2^{\expSize-1}$, and $2^{\expSize-1}$.

\subsection{High-level Description of F2P}\label{sec:F2P_overview}
In standard Floating-point representation~\cite{IEEE754_lec_nots}, the {\em value} of the exponent floats, but the {\em size} of the exponent field is fixed. F2P adds flexibility to FP by letting the {\em size} of the exponent field ``float'' as well.
To encode the actual size of the exponent field, we introduce the {\em hyper-exp} field, which specifies the number of bits in the exponent field. As the number of bits required to represent a number is only logarithmic in its value, the hyper-exp field can be tiny (e.g., only two bits). 

While designing F2P, we aim to minimize the hardware changes required to extend an FP system to handle F2P numbers. 
Accordingly, given the mantissa value $\mantVal$ and the exponent value $\expVal$, the 
value $\cntrVal(\expVal, \mantVal)$ inferred by an F2P system is identical to the value inferred by an FP system, namely, the value detailed in~\eqref{eq:FP_val}. 
Furthermore, the encoding of the mantissa in an F2P system is identical to the encoding in an FP system, namely, the value $\mantVal(\mantVec)$ detailed in~\eqref{eq:def_expVal_mantVal}. 

However, F2P's exponent encoding differs from that of FP. F2P's novel exponent encoding utilizes a scalable exponent field to fine-tune the range of numbers with increased accuracy. 
In particular, F2P's exponent encoding uses the following design principles.

\begin{table}[b!]
    \footnotesize
    \centering
    \caption{Two possible ways to interpret an exponent vector $\expVec$ of up to two bits. The size of the vector is $\expSize$. `-' represents an empty (zero-size) vector.
    SR, and LR represents the interpretation of $\expVec$ for the Small Reals, and Large Reals flavors of F2P.}
    \label{tab:F2P_intuition}
    \begin{tabular}{|c|c|c|c|}
        \hline
        $\expSize$ & $\expVec$ & \expValSr & \expValLr \\ \hline
        0 & - & 0 & 0 \\ \hline
        1 & 0 & 1 & -1 \\ 
        1 & 1 & 2 & -2 \\ \hline
        2 & 00 & 3 & -3 \\ 
        2 & 01 & 4 & -4 \\ 
        2 & 10 & 5 & -5 \\ 
        2 & 11 & 6 & -6 \\ \hline
    \end{tabular}
\end{table}

\noindent{\em The exponent-field scaling principle}: 
F2P's scalable mantissa field allows the designer to select a range of improved accuracy. Tab.~\ref{tab:F2P_intuition} presents all the possible binary vectors with at most two bits. For each such vector, consider two ways to interpret the vector $\expVec$: (i) using an increasing order. This way, smaller numbers use a smaller exponent field, thus leaving more bits for the mantissa field of small numbers. This interpretation boosts the accuracy when representing small real values; we dub this option Small Reals (SR). (ii) Use a decreasing order by adding the minus sign, as suggested in the rightmost colon. This way, {\em larger} numbers use smaller exponent fields. Consequently, the mantissa field of a larger number is longer, implying improved accuracy for large reals; we dub this option Large Reals (LR). When using either SR or LR, one can adjust the counting range by fine-tuning the bias $\bias$. Specifically, we will determine the bias $\bias$ using the symmetrical power principal detailed in Sec.~\ref{sec:FP}. 

\begin{table*}[t!]
    \small
    \caption{\label{tab:F2P_examples}Examples of 6-bit numbers of different F2P flavors with $\hyperSize=2$. The hyper-exp $\hyperVec$, exponent $\expVec$, and mantissa $\mantVec$ are typed in purple, blue, and black, respectively.}
    \label{tab:f2p_examples}
    \begin{tabular}{|p{0.9cm}|p{0.5cm}|l|l|l|l|}
        \cline{3-6}
        \multicolumn{2}{l|}{}& \begin{tabular}[c]{@{}l@{}}
        \rule[1pt]{0pt}{2.0ex}    
        \ffpSrTwo, \bias= -8, E=$\val (\expVec)$ \end{tabular} & 
        \begin{tabular}[c]{@{}l@{}}
        \rule[1pt]{0pt}{2.0ex}    
        \ffpLrTwo, \bias=7, E=-$\val (\expVec)$\end{tabular} & 
        \begin{tabular}[c]{@{}l@{}}
        \rule[1pt]{0pt}{2.0ex}    
        \ffpSiTwo, \bias=3, E=$\val (\expVec)$\end{tabular} & 
        \begin{tabular}[c]{@{}l@{}}
        \rule[1pt]{0pt}{2.0ex}    
        \ffpLiTwo, \bias=14,  E= -$\val (\expVec)$
        \rule[1pt]{0pt}{2.0ex}    
        \end{tabular} 
        % \rule[1pt]{0pt}{2.0ex}    
        \rule[-5pt]{0pt}{2.0ex}    
        \\ \hline
       \multicolumn{1}{|l|}{$\cntrVec$} & \multicolumn{1}{p{3.0mm}|}{$\val (\expVec)$} & $\cntrVal$ & $\cntrVal$ & $\cntrVal$ & $\cntrVal$ 
       \\ \hline\hline
       \textcolor{purple}{00}{\blue}0000  
       & 0  & $0.0000 \cdot 2^{-7}=0$&  $1.0000 \cdot 2^7$=128 &  $0.0000 \cdot 2^4$=0 &  $1.0000 \cdot 2^{14}$=16384 
       \rule[2ex]{0pt}{0.1ex}  
       \\
       \textcolor{purple}{00}{\blue}0001  & 0  & $0.0001 \cdot 2^{-7}$=1/2048  & $1.0001 \cdot 2^7$=136 &  $0.0001 \cdot 2^{-7}$=1 & $1.0001 \cdot 2^{14}$=17408 \\
       \textcolor{purple}{00}{\blue}1111  & 0  & $0.1111 \cdot 2^{-7}$=15/2048 &  $1.1111 \cdot2^7$=248 &  $0.1111 \cdot 2^4$=15 &  $1.1111 \cdot 2^{14}$=31744 \\ \hline
       
       \textcolor{purple}{01}{\blue 0}000 & 1  & $1.000 \cdot2^{-7}$=16/2048  & $1.000 \cdot 2^6$=64   &  $1.000 \cdot2^4$=16 & $1.000 \cdot2^{13}$=8192   
       \rule[2ex]{0pt}{0.1ex}  
       \\ 
       \textcolor{purple}{01}{\blue 0}001 & 1  & $1.001 \cdot2^{-7}$=18/2048  &  $1.001 \cdot2^6=72$        &  $1.001 \cdot2^4$=18 &$1.001 \cdot2^{13}$=9216  \\ 
       \textcolor{purple}{01}{\blue 0}111 & 1  & $1.111 \cdot2^{-7}$=30/2048  & $1.111 \cdot2^6$=120  & $1.111 \cdot2^4$=30  & $1.111 \cdot2^{13}$=15360  \\ \hline
       \textcolor{purple}{01}{\blue 1}000 & 2  & $1.000 \cdot2^{-6}$=32/2048  &  $1.000 \cdot2^5$=32     & $1.000 \cdot2^5$=32  & $1.000 \cdot2^{12}$=4096   
       \rule[2ex]{0pt}{0.1ex}  
       \\\hline
       \textcolor{purple}{11}{\blue 110}0 & 13 & $1.0 \cdot2^5$=32 & $1.0 \cdot2^{-6}$=1/64  &  $1.0 \cdot2^{16}$=65536     &  $1.0 \cdot2^1$=2 
       \rule[2ex]{0pt}{0.1ex}  
       \\ \hline
       \textcolor{purple}{11}{\blue 111}0 &14  & $1.0 \cdot2^6$=64 &  $0.0 \cdot2^{-6}$=0 &  $1.0 \cdot2^{17}$=131072 & $0.0 \cdot2^1$=0 \\ 
       \textcolor{purple}{11}{\blue 111}1 &14  & $1.1 \cdot2^6$=96 &  $0.1 \cdot2^{-6}$=1/128 &  $1.1 \cdot2^{17}$=196608    &  $0.1 \cdot2^1$=1 \\ \hline
\end{tabular}
\end{table*}

\begin{table}[b]
    \centering
    \footnotesize
    \caption{Exponent and bias values of F2P's flavors}\label{tab:F2P_flavors}
    \begin{tabular}{|c|c|c|}
        \hline
        F2P flavor  &  \expVal & \bias \\ \hline
        \ffpSr      & $\val (\expVec)$  & $ -0.5\left(\valMax+1 \right)$
        \\ \hline
        \ffpLr      & $-\val (\expVec)$  & $0.5\left(\valMax-1\right)$ \\ \hline
        \ffpSi      & $\val (\expVec)$  & $\rule{0pt}{10pt}\cntrSize - \hyperSize - 1$ \\ \hline     
        \ffpLi      & $-\val (\expVec)$  & $\rule{0pt}{10pt}\cntrSize - \hyperSize - 2^\hyperSize + \valMax - 1$ \\ \hline
    \end{tabular}
\end{table}

\subsection{F2P's Exponent Encoding}
We now detail F2P's exponent encoding.
F2P$^\hyperSize$ denotes an F2P system with $\hyperSize$ hyper-exp bits. Denote the hyper-exp vector by $\hyperVec = \hyperBit_{\hyperSize-1} \dots \hyperBit_0$. The hyper-exp field uses an
unsigned binary encoding, namely $\expSize (\hyperVec)=\sum_0^{\hyperSize-1} 2^{\hyperBit_i}$. 
Consequently, the number of bits in the exponent field varies between 0 and $\expMaxSize = 2^\hyperSize-1$. 

To effectively utilize every bit in the exponent field, we decode an $\expSize$-bits exponent vector while considering not only the $\expSize$-bits vector $\expVec$ itself but also all the values that could be encoded using shorter vectors. 
An $i$-bit binary vector can encode $2^i$ distinct values. It follows that the number of values that can be encoded using binary vectors of sizes $0, \dots, \expSize-1$ is $\sum_{i=0}^{\expSize-1} 2^i$. 
Given an $\expSize$-bits exponent vector $\expVec = \expBit_{\expSize-1} \dots \expBit_{0}$ we therefore define the corresponding exponent value 
\begin{align}\label{eq:def_expVal_of_expVec}
    \val (\expVec) = \sum_{i=0}^{\expSize-1} 2^i +
    \sum_{i=0}^{\expSize-1} \expBit_i 2^i 
    = 
    \sum_{i=0}^{\expSize-1} (1+\expBit_i) 2^i.
\end{align}

$\val (\expVec)$ is maximized when $\expVec$ is a vector of $2^{\hyperSize}-1$ ones. 
Hence, the number of distinct values that can be encoded by exponent vectors of up to $2^\hyperSize-1$ bits is
\begin{align}\label{eq:valMax}
    \valMax = \sum\limits_{i=0}^{2^\hyperSize-1} 2^i = 2^{\left(2^\hyperSize\right)}-1.
    \end{align}

Using the design principles detailed in~\ref{sec:F2P_overview}, we now explore different flavors of F2P, customized to the needs of different applications.

\paragraph*{Focus on small reals} We denote F2P's variant, which focuses on accurately representing small real numbers \ffpSr. 
Based on the exponent field scaling principle, the
exponent's value should increase with the size of the exponent field. Hence, we set $\expVal_{\sr} (\expVec) = \val(\expVec)$.
By the symmetrical power principle, we set the bias value $\bias_{\sr} = -\left(\frac{\valMax+1}{2}\right)$ (note that by~\eqref{eq:valMax}, $\valMax$ is odd).

Tab.~\ref{tab:F2P_examples} exemplifies several values of a 6-bit \ffpSr, where the size of the hyper-exp field is $\hyperSize=2$. By~\eqref{eq:valMax}, $\valMax=15$. Consequently, the bias is $\bias_{\sr} = -(15+1)/2 = -8$ 

\paragraph*{Focus on large reals} We denote F2P's variant, which focuses on large real numbers \ffpLr. 
Based on the exponent field scaling principle, the exponent field's size should decrease with the value of the number. Hence, we set $\expVal_{\lr} (\expVec) = -\val(\expVec)$. By the symmetrical power principle, we set the bias value $\bias_{\lr} = \frac{\valMax-1}{2}$. 
  Tab.~\ref{tab:F2P_examples} depicts several values of a 6-bit \ffpLr, with a 6-bits hyper-exp field. Consequently, $\valMax=15$ and $\bias_{\lr} = (15-1)/2 = 7$.

\subsection{Representing Integers}

We now customize F2P to represent only integers.

To represent only integers, the smallest positive value that the system represents should be 1. 
By~\eqref{eq:FP_val}, this is equivalent to choosing a bias value satisfying 
\begin{align}\label{eq:int_cond}
    \mantVal_{min} \cdot 2^{\expMinVal+\bias+1} =1,    
\end{align}
where $\mantVal_{min}$ is the smallest mantissa's value. 
To calculate $\mantVal_{min}$, observe that 
the smallest positive value is obtained when the mantissa vector is $\mantVec = 0 \dots 01$. Then, by~\eqref{eq:def_expVal_mantVal}, 
\begin{align}\label{eq:mantMinVal}
    \mantVal_{min} = 2^{-\mantSize},
\end{align}
where $\mantSize$ is the size of the mantissa field, to be determined later.
Assigning~\eqref{eq:mantMinVal} in~\eqref{eq:int_cond}, we obtain
\begin{align}\label{eq:int_cond_by_mantSize}
    \bias = \mantSize - \expMinVal - 1.
\end{align}
Using~\eqref{eq:int_cond_by_mantSize}, we now develop concrete flavors of F2P that represent only integers. 

\paragraph*{Focus on small integers} We dub this F2P's variant \ffpSi. 
Based on the exponent-field scaling principle, we define $\expVal_{\si} (\expVec) = \val(\expVec)$ and 
$\expMinVal=0$. 
To accurately represent small integers, the smallest numbers should use the smallest possible exponent size, namely, zero. 
The size of the mantissa field is the leftover of the number's size after deducing the hyper-exp, namely, $\mantSize = \cntrSize - \hyperSize$
Assigning $\expMinVal=0$ and $\mantSize = \cntrSize - \hyperSize$ in~\eqref{eq:int_cond_by_mantSize}, we obtain 
$\bias_{\si} = \cntrSize - \hyperSize - 1.$    
Tab.~\ref{tab:F2P_examples} depicts several values of a 6-bit \ffpSi. The size of the hyper-exp field is $\hyperSize=2$. Correspondingly, $\valMax=15$ and $\bias_{\si} = 6-2-1 = 3$ 

\paragraph*{Focus on large integers} We dub this F2P's variant \ffpLi. 
Based on the exponent-field scaling principle, we define $\expVal_{\li} (\expVec) = -\val(\expVec)$. It follows that 
\begin{align}\label{eq:expMinVal_of_li}
    \expMinVal = -(\valMax-1).
\end{align}
To analyze the size of the mantissa field $\mantSize$ in~\eqref{eq:int_cond_by_mantSize}, consider again the exponent-field scaling principle. To accurately represent large numbers, the smallest numbers should use the largest exponent field, namely $\expMaxSize = 2^{\hyperSize}-1$. The size of the mantissa field is the leftover of the number's size after deducing the hyper-exp and the (largest possible) exponent field, namely, 
\begin{align}\label{eq:mantMinSize} 
    \mantSize = \cntrSize - \hyperSize - 2^{\hyperSize}+1
\end{align}
Assigning the expressions for $\expMinVal$~\eqref{eq:expMinVal_of_li} and for $\mantSize$~\eqref{eq:mantMinSize} in~\eqref{eq:int_cond_by_mantSize}, we obtain 
\begin{align}
    \bias_{\li} = \cntrSize - \hyperSize - 2^\hyperSize + \valMax-1.    
\end{align}
Tab.~\ref{tab:F2P_examples} depicts several values of a 6-bit \ffpLi. The size of the hyper-exp field is $\hyperSize=2$. Correspondingly, $\valMax=15$ and $\bias_{\li} = 6-2-2^2 + 15-1 = 14$.

\noindent The exponent value $\expVal$ and the bias $\bias$ of all the flavors of F2P considered in this paper are summarized in Tab.~\ref{tab:F2P_flavors}. 

\section{Evaluation}\label{sec:sim}
\noindent In this section, we compare the performance of the F2P number representation with state-of-the-art solutions. We focus on two domains, namely, approximate counters (Sec.~\ref{sec:eval:cntrs}) and neural network quantization (Sec.~\ref{sec:eval:quant}).
In our evaluation, we consider the Mean Square Error (MSE), defined as follows: Given exact values $V_1, \dots, V_S$ and the corresponding approximated values $C_1, 
\dots, C_S$, the MSE is $\frac{1}{S} \sum_{i=1}^{S} (C_i-V_i)^2$.
% exact value $e_i$ and estimated value
% errors $\err_1, \dots, \err_S$, the MSE is $\frac{1}{S} \sum_{i=1}^{S} \err_i$.

\subsection{Using F2P as Approximate Counters}\label{sec:eval:cntrs}

An array of counters, counting numbers belonging to heterogeneous ranges, is a fundamental building block in numerous
applications, including computer 
vision, traffic engineering, intrusion detection, congestion control, and natural language processing~\cite{Pixel_cntrs17, Sead_ToN, Sketch_survey_arxiv20, A_sketch16}.

We compare F2P's performance to state-of-the-art solutions as follows.  
Consider a perfect counter that is incremented by one each time (e.g., to count packet arrivals). Let $C_i$ denote the value of the approximate counter (F2P, or one of our benchmarks, to be detailed shortly) after the $i$-th increment of the perfect counter.
The {\em on-arrival model}~\cite{on_arrival} measures the error upon each such increment. After $S$ such increments, the MSE is $\frac{1}{S} \sum_{i=1}^{S} (C_i-i)^2$.

For this experiment, we consider F2P's LI (Large Integers) flavor, as we need to count integers, with a focus on the accuracy when measuring large integers, which represent large flows~\cite{Sketch_survey_arxiv20}.
We consider the following benchmark: Morris~\cite{Morris1978}, CEDAR~\cite{CEDAR_Tsidon}, and dynamic SEAD~\cite{Sead_ToN}. We vary the counter-width $\cntrSize$. For Morris and CEDAR, we find, through a binary search, the configuration parameters that minimize the error while still reaching the maximal number that \ffpLrTwo\ reaches for the same counter-width. Dynamic SEAD does not have configuration parameters.
The results of this experiment are shown in Tab.~\ref{tab:single_cntr}.
To make a meaningful comparison, we normalize the results of each counter size -- namely, each row in the table -- w.r.t. the lowest value obtained in that row.

The results show that F2P reaches the lowest MSE across the board. The worst accuracy is obtained by dynamic SEAD, due to its wasteful exponent's unary representation, which shrinks the mantissa field, and consequently compromises accuracy.

\begin{table}[]
    \centering
    \footnotesize
    \caption{Comparison of the accuracy of approximate counter solutions. For each counter width (captured by a row in the table), the values are normalized w.r.t. the smallest value for this counter size. }
        \label{tab:single_cntr}
    \begin{tabular}{|c|c|c|c|c|}
        \hline
        \rule[0.5pt]{0pt}{2.0ex}    
        \# bits & \ffpLiTwo & CEDAR & Morris & SEAD  
        \rule[-3pt]{0pt}{2.0ex}    
        \\ \hline
            8 & \green{\textbf{1.00}} & 1.71 & 1.80 & 124.55 \\
            10 & \green{\textbf{1.00}} & 1.75 & 1.80 & 468.49 \\
            12 & \green{\textbf{1.00}} & 2.05 & 1.94 & 1687.06 \\
            14 & \green{\textbf{1.00}} & 2.05 & 1.67 & 6538.85 \\
            16 & \green{\textbf{1.00}} & 1.77 & 2.04 & 31420.84 \\
        \hline
        \end{tabular}
\end{table}

% \begin{table}[]
%     \centering
%     \caption{Comparison of the accuracy of a Count Min Sketch with different estimators.}
%     \label{tab:Caida}
%     \begin{tabular}{|c|c|c|c|c|}
%         \hline
%         \# bits & \ffpLiTwo & CEDAR & Morris & SEAD  \\ \hline
%         \hline
%         \end{tabular}
% \end{table}

\subsection{Using F2P for Neural Networks Quantization}
\label{sec:eval:quant}

We now study the performance of F2P for neural network quantization. We focus on the quantization error and, therefore, consider the min-max approach, which generates only quantization errors while nullifying other errors~\cite{Quant_white_paper}.
Given a vector $V$, let $V^{F}$ denote its counterpart quantized vector that belongs to the number system $F$.
Define the scaling factor $s = \frac{\max V - \min V}{F_{max} - F_{min}}$, where $F_{max}$ and $F_{min}$ are the maximum and minimum values that can be represented by $F$. 
Then, $V^F = s \cdot \nint*{\frac{V}{s}}$, where 
$\nint*{}$ denotes the round-to-nearest operation.
The {\em quantization error} of the $i$-the element in the vector is $\textrm{err}_i = \abs{v_i - v^F_i}$~\cite{FP8_quant}. 
We consider the Mean Square error (MSE) 
which is indicative of the final loss in a neural network~\cite{up_or_dwn}.

%    // ABSOLUTE
\begin{table*}[]
    \scriptsize
 %   // ABSOLUTE
    \caption{MSE of the quantization error for different number systems. Values at each setting (row) are normalized w.r.t. the smallest value in that row. Values within 1\% of the minimum for that setting are green-colored.}
    \label{tab:NN_MSS}
    % \centering
    \begin{tabular}{|c||c|c|c|c|c|c|c|c||c|c|c|c|c|c|}
    \hline
    \rule[0.5pt]{0pt}{2.0ex}    
    Model & \ffpSrOne & \ffpSrTwo & \ffpLrOne & \ffpLrTwo & \ffpSiOne & \ffpSiTwo & \ffpLiOne & \ffpLiTwo & INT8 & SEAD & 5ME2 &4Me3 & 3ME4 & 2ME5
    \rule[-3pt]{0pt}{2.0ex}    
    \\ 
    \hline
		% Uniform & 2.3 & 156.9 & \green{\textbf{1.0}} & 3.2 & 2.3 & 156.9 & \green{\textbf{1.0}} & 3.2 & \green{\textbf{1.0}} & 49.3 & 2.3 & 9.3 & 37.7 & 156.9 \\ 
		% Normal & 2.4 & 148.4 & \green{\textbf{1.0}} & 4.0 & 2.4 & 148.4 & \green{\textbf{1.0}} & 4.0 & \green{\textbf{1.0}} & 6.7 & 2.4 & 9.6 & 38.2 & 148.4 \\ 
		% t, $\nu=5$ & 1.1 & 73.6 & \green{\textbf{1.0}} & 4.0 & 1.1 & 73.6 & \green{\textbf{1.0}} & 4.0 & \green{\textbf{1.0}} & 2.6 & 1.1 & 4.2 & 17.2 & 73.6 \\ 
		% t, $\nu=8$ & 3.6 & 211.4 & \green{\textbf{1.0}} & 4.0 & 3.6 & 211.4 & \green{\textbf{1.0}} & 4.0 & \green{\textbf{1.0}} & 7.0 & 3.6 & 14.2 & 56.0 & 211.4 \\ 
		Resnet18 & \green{\textbf{1.0}} & 32.9 & 2.4 & 2.3 & \green{\textbf{1.0}} & 32.9 & 2.4 & 2.3 & 2.4 & 1.8 & \green{\textbf{1.0}} & 2.2 & 8.4 & 32.9 \\ 
		Resnet50 & 13.2 & 37.6 & 54.6 & 2.3 & 13.2 & 37.6 & 54.6 & 2.3 & 55.0 & 5.2 & 13.2 & \green{\textbf{1.0}} & 3.1 & 37.6 \\ 
		MNet\_V2 & 1.5 & 32.8 & 26.7 & 20.8 & 1.5 & 32.8 & 26.7 & 20.8 & 20.8 & \green{\textbf{1.0}} & 1.5 & 6.1 & 1.3 & 32.8 \\ 
		MNet\_V3 & 17.9 & 2.0 & 2.3e+04 & 1.3 & 17.9 & 2.0 & 2.3e+04 & 1.3 & 2.3e+04 & 1147.8 & 17.9 & 3.8 & \green{\textbf{1.0}} & 2.3 \\ 

        \hline
    \end{tabular}

\vspace{0.5 cm}

    \begin{tabular}{|c||c|c|c|c|c|c|c|c||c|c|c|c|}
    \hline
    \rule[0.5pt]{0pt}{2.0ex}    
    Model & \ffpSrOne & \ffpSrTwo & \ffpLrOne & \ffpLrTwo & \ffpSiOne & \ffpSiTwo & \ffpLiOne & \ffpLiTwo & INT16 & SEAD & FP16 & BF16
    \rule[-3pt]{0pt}{2.0ex}    
    \\ 
    \hline
		% Uniform & 2.3 & 146.3 & \green{\textbf{1.0}} & 3.2 & 2.3 & 146.3 & \green{\textbf{1.0}} & 3.2 & \green{\textbf{1.0}} & 2.9e+05 & 146.3 & 9384.9 \\ 
		% Normal & 2.4 & 156.6 & \green{\textbf{1.0}} & 4.0 & 2.4 & 156.6 & \green{\textbf{1.0}} & 4.0 & \green{\textbf{1.0}} & 1710.3 & 156.6 & 1.0e+04 \\ 
		% t, $\nu=5$ & 1.1 & 67.8 & \green{\textbf{1.0}} & 4.0 & 1.1 & 67.8 & \green{\textbf{1.0}} & 4.0 & \green{\textbf{1.0}} & 93.7 & 67.8 & 4333.3 \\ 
		% t, $\nu=8$ & 3.6 & 229.2 & \green{\textbf{1.0}} & 4.0 & 3.6 & 229.2 & \green{\textbf{1.0}} & 4.0 & \green{\textbf{1.0}} & 1034.4 & 229.2 & 1.5e+04 \\ 
		Resnet18 & \green{\textbf{1.0}} & 35.0 & 2.4 & 2.5 & \green{\textbf{1.0}} & 35.0 & 2.4 & 2.5 & 2.4 & 1201.7 & 35.0 & 2618.8 \\ 
		Resnet50 & 5.5 & 14.1 & 21.7 & \green{\textbf{1.0}} & 5.5 & 14.1 & 21.7 & \green{\textbf{1.0}} & 21.6 & 67.9 & 14.2 & 1294.2 \\ 
		MNet\_V2 & 2.0 & 32.7 & 8.3 & 8.2 & 2.0 & 32.7 & 8.3 & 8.2 & 8.3 & \green{\textbf{1.0}} & 32.7 & 1926.7 \\ 
		MNet\_V3 & 46.0 & \green{\textbf{1.0}} & 2426.9 & 2.6 & 46.0 & \green{\textbf{1.0}} & 2426.9 & 2.6 & 2432.6 & 21.9 & 4.8 & 567.0 \\ 
        
                \hline
    \end{tabular}
    
\vspace{0.5 cm}

    \begin{tabular}{|c||c|c|c|c|c|c|c|c||c|c|c|c|c|} % {|M{0.12\columnwidth}
    % \begin{tabular}{|c||c|c|c|c|c|c|c|c||c|c|}
    % \begin{tabular}{|M{0.17\columnwidth}|m{0.71\columnwidth}|}
    \hline
    \rule[0.5pt]{0pt}{2.0ex}    
    Model & \ffpSrOne & \ffpSrTwo & \ffpLrOne & \ffpLrTwo & \ffpSiOne & \ffpSiTwo & \ffpLiOne & \ffpLiTwo & INT19 & SEAD & TP19
    \rule[-3pt]{0pt}{2.0ex}    
    \\ 
    \hline
		% Uniform & 2.3 & 146.3 & \green{\textbf{1.0}} & 3.2 & 2.3 & 146.3 & \green{\textbf{1.0}} & 3.2 & \green{\textbf{1.0}} & 1.1e+07 & 146.3 \\ 
		% Normal & 2.5 & 156.6 & \green{\textbf{1.0}} & 4.0 & 2.5 & 156.6 & \green{\textbf{1.0}} & 4.0 & \green{\textbf{1.0}} & 2.4e+04 & 156.6 \\ 
		% t, $\nu=5$ & 1.1 & 67.4 & \green{\textbf{1.0}} & 4.0 & 1.1 & 67.4 & \green{\textbf{1.0}} & 4.0 & \green{\textbf{1.0}} & 563.6 & 67.4 \\ 
		% t, $\nu=8$ & 3.6 & 229.2 & \green{\textbf{1.0}} & 4.0 & 3.6 & 229.2 & \green{\textbf{1.0}} & 4.0 & \green{\textbf{1.0}} & 9840.2 & 229.2 \\ 
		Resnet18 & \green{\textbf{1.0}} & 38.3 & 2.5 & 2.4 & \green{\textbf{1.0}} & 38.3 & 2.5 & 2.4 & 2.6 & 6.2e+04 & 38.3 \\ 
		Resnet50 & 7.9 & 4.2 & 29.7 & \green{\textbf{1.0}} & 7.9 & 4.2 & 29.7 & \green{\textbf{1.0}} & 29.9 & 2628.2 & 4.2 \\ 
		MNet\_V2 & \green{\textbf{1.0}} & 15.8 & 4.0 & 4.0 & \green{\textbf{1.0}} & 15.8 & 4.0 & 4.0 & 4.0 & 2.2 & 15.8 \\ 
		MNet\_V3 & 133.5 & \green{\textbf{1.0}} & 358.6 & 2.6 & 133.5 & \green{\textbf{1.0}} & 358.6 & 2.6 & 357.4 & 11.5 & 3.4 \\

    \hline
    \end{tabular}
\end{table*}

We consider the weights in the pre-trained models Resnet18, Resnet50, MobileNet\_V2, and MobileNet\_V3~\cite{PyTorch_Models}. 
We consider the following FP formats. For eight bits, there exists no FP standard, and therefore we consider 2M5E, 3M4E, 4M3E, and 5M2E.
% , as in~\cite{FP8_quant}. 
For 16-bits, we consider FP16 (10M5E) and BF16 (7M8E)~\cite{BFloatBF16GoogleCloud}. We also consider the 19-bit TF32 (10M8E)~\cite{TensorFloat32Nvidia}. Finally, we consider for all bit-widths the corresponding integer representations and dynamic SEAD~\cite{Sead_ToN}.
The results are detailed in Tab.~\ref{tab:NN_MSS}. To ease the comparison, the values for each setting -- namely, each row in the table -- are normalized w.r.t. the lowest value in that row.

The results show that the F2P flavors obtain a lower MSE than standard 16- and 19-bit formats for most tested models. Considering the 8-bit formats, however, F2P typically does worse than the best standard-format alternative. Intuitively, paying the overhead of the hyper-exp field introduced by F2P becomes more beneficial when the bit count is not too small.

\section{Conclusion}\label{sec:conclusion}
\noindent In this work, we present F2P - a novel number system that provides a robust counting range and higher accuracy on a select range. 
We suggest different flavors of F2P that differ on the selected range and focus (e.g., integers, large numbers, or small numbers). We then show that such added flexibility improves performance on realistic scenarios such as network measurement and federated learning. Namely, our work improves on Morris counters~\cite{Morris1978}, CEDAR~\cite{CEDAR_Tsidon}, and SEAD~\cite{Sead_ToN} as well as on popular and widely available fixed-length numbers such as INT8, and short floating points numbers (e.g., FP16, BF16~\cite{BFloatBF16GoogleCloud}, TensorFloat32~\cite{TensorFloat32Nvidia}.

F2P uses the same notions as regular floating point numbers, which are extensively deployed. The main difference between the two is that F2P's partition into sign and mantissa is flexible. This flexibility is essential when using short numbers that are gaining popularity due to recent trends. In the future, we will pursue a full hardware implementation of F2P. Such an endeavor looks promising as F2P calculations are nearly identical to floating point arithmetic, which admits to efficient implementations. 

%Further, our design keeps a low delay, space, and power overheads.
%Admittedly, the preliminary results reported here discard the configuration dilemmas and nitty-gritty details that dwell in a practical implementation. \gil{are you sure you want that sentence in the conclusion? it feels like you are just helping a negative review give justification. }
%In future work, we plan to develop an arithmetic unit that supports F2P and study the feasibility of the approach for hardware implementation.  \gil{Check up this sentence edit if you think it is bad/want to say something else. I didn't want to write anything apologetic like in the future we'll run more workloads because it is like saying we didn't evaluate enough. 
%By the way, I think we can do a hardware implementation, I have a research group in mind that might be interested in leading such an endeavor.  At least its technical hardware design aspects. }

%into existing applications and evaluate its performance using additional real-world workloads.

% sketching mechanisms and in training-aware quantization and post-training quantization.
% bla bla

\bibliographystyle{IEEEtran}
\bibliography{Refs}

\begin{thebibliography}{10}
\providecommand{\url}[1]{#1}
\csname url@samestyle\endcsname
\providecommand{\newblock}{\relax}
\providecommand{\bibinfo}[2]{#2}
\providecommand{\BIBentrySTDinterwordspacing}{\spaceskip=0pt\relax}
\providecommand{\BIBentryALTinterwordstretchfactor}{4}
\providecommand{\BIBentryALTinterwordspacing}{\spaceskip=\fontdimen2\font plus
\BIBentryALTinterwordstretchfactor\fontdimen3\font minus
  \fontdimen4\font\relax}
\providecommand{\BIBforeignlanguage}[2]{{%
\expandafter\ifx\csname l@#1\endcsname\relax
\typeout{** WARNING: IEEEtran.bst: No hyphenation pattern has been}%
\typeout{** loaded for the language `#1'. Using the pattern for}%
\typeout{** the default language instead.}%
\else
\language=\csname l@#1\endcsname
\fi
#2}}
\providecommand{\BIBdecl}{\relax}
\BIBdecl

\bibitem{Quant_white_paper}
M.~Nagel, M.~Fournarakis, R.~A. Amjad, Y.~Bondarenko, M.~Van~Baalen, and
  T.~Blankevoort, ``A white paper on neural network quantization,'' \emph{arXiv
  preprint arXiv:2106.08295}, 2021.

\bibitem{FP8_quant}
A.~Kuzmin, M.~Van~Baalen, Y.~Ren, M.~Nagel, J.~Peters, and T.~Blankevoort,
  ``Fp8 quantization: The power of the exponent,'' \emph{Advances in Neural
  Information Processing Systems}, vol.~35, pp. 14\,651--14\,662, 2022.

\bibitem{Sketch_survey_arxiv20}
S.~Li, L.~Luo, D.~Guo, Q.~Zhang, and P.~Fu, ``A survey of sketches in traffic
  measurement: Design, optimization, application and implementation,''
  \emph{arXiv preprint arXiv:2012.07214}, 2020.

\bibitem{Pixel_cntrs17}
F.~M. Della~Rocca, T.~Al~Abbas, N.~A. Dutton, and R.~K. Henderson, ``A high
  dynamic range spad pixel for time of flight imaging,'' in \emph{2017 IEEE
  SENSORS}.\hskip 1em plus 0.5em minus 0.4em\relax IEEE, 2017, pp. 1--3.

\bibitem{BFloatBF16GoogleCloud}
\BIBentryALTinterwordspacing
``The bfloat16 numerical format.'' [Online]. Available:
  \url{https://cloud.google.com/tpu/docs/bfloat16}
\BIBentrySTDinterwordspacing

\bibitem{TensorFloat32Nvidia}
\BIBentryALTinterwordspacing
``{TensorFloat-32 in the A100 GPU Accelerates AI Training, HPC up to 20x}.''
  [Online]. Available:
  \url{https://blogs.nvidia.com/blog/tensorfloat-32-precision-format/}
\BIBentrySTDinterwordspacing

\bibitem{Sead_ToN}
X.~Liu, Y.~Xu, P.~Liu, T.~Yang, J.~Xu, L.~Wang, G.~Xie, X.~Li, and S.~Uhlig,
  ``Sead counter: Self-adaptive counters with different counting ranges,''
  \emph{IEEE/ACM Transactions on Networking}, vol.~30, no.~1, pp. 90--106,
  2021.

\bibitem{CEDAR_Tsidon}
E.~Tsidon, I.~Hanniel, and I.~Keslassy, ``Estimators also need shared values to
  grow together,'' in \emph{INFOCOM}.\hskip 1em plus 0.5em minus 0.4em\relax
  IEEE, 2012, pp. 1889--1897.

\bibitem{Morris1978}
R.~Morris, ``Counting large numbers of events in small registers,''
  \emph{Communications of the ACM}, vol.~21, no.~10, pp. 840--842, 1978.

\bibitem{IEEE754_lec_nots}
W.~Kahan, ``{IEEE} standard 754 for binary floating-point arithmetic,''
  \emph{Lecture Notes on the Status of IEEE}, vol. 754, no. 94720-1776, p.~11,
  1996.

\bibitem{A_sketch16}
P.~Roy, A.~Khan, and G.~Alonso, ``Augmented sketch: Faster and more accurate
  stream processing,'' in \emph{Proceedings of the 2016 International
  Conference on Management of Data}, 2016, pp. 1449--1463.

\bibitem{on_arrival}
R.~B. Basat, X.~Chen, G.~Einziger, R.~Friedman, and Y.~Kassner, ``Randomized
  admission policy for efficient top-k, frequency, and volume estimation,''
  \emph{IEEE/ACM Transactions on Networking}, vol.~27, no.~4, pp. 1432--1445,
  2019.

\bibitem{up_or_dwn}
M.~Nagel, R.~A. Amjad, M.~Van~Baalen, C.~Louizos, and T.~Blankevoort, ``Up or
  down? adaptive rounding for post-training quantization,'' in
  \emph{International Conference on Machine Learning}.\hskip 1em plus 0.5em
  minus 0.4em\relax PMLR, 2020, pp. 7197--7206.

\bibitem{PyTorch_Models}
\BIBentryALTinterwordspacing
``{PyTorch Models}.'' [Online]. Available:
  \url{https://pytorch.org/vision/main/models.html}
\BIBentrySTDinterwordspacing

\end{thebibliography}
\end{document}